\documentclass[final,5p,times,twocolumn]{elsarticle}

\usepackage{graphicx}
\usepackage{dcolumn}
\usepackage{bm}
\usepackage{amsfonts}
\usepackage{latexsym}
\usepackage{amssymb}
\usepackage{verbatim}
\usepackage{amsthm}
\usepackage{amsmath}

\begin{document}

\begin{frontmatter}



\title{
  An application of the measurement of expectation values \\
  for the photon annihilation and creation operators
}


\author{Kouji Nakamura}
\ead{kouji.nakamura@nao.ac.jp}

\author{Masa-Katsu Fujimoto}
\ead{fujimoto.masa-katsu@nao.ac.jp}

\address{
  Gravitational-Wave Project Office,
  Optical and Infrared Astronomy Division,
  National Astronomical Observatory of Japan,
  Mitaka, Tokyo 181-8588, Japan
}

\begin{abstract}
Motivated by the readout scheme in interferometric gravitational-wave
detectors, we consider the device which measures the expectation
value of the photon annihilation and creation operators for output
optical field from the main interferometer.
As the result, the eight-port homodyne detection is rediscovered as
such a device.
We evaluate the noise spectral density in this measurement.
We also briefly discuss on the application of our results to the
readout scheme of gravitational-wave detectors.
We call this measurement scheme to measure these expectation values as
``{\it double balanced homodyne detection.}''
\end{abstract}


\end{frontmatter}


\section{Introduction}
\label{sec:Introduction}


In quantum theory, an ``observable'' is represented by a self-adjoint
operator acting on the Hilbert space as an axiom.
Besides of the origin of the terminology of ``observable,'' it is
sometime discussed which variable is measurable in quantum theory.
The photon phase difference is a typical example and there are many
literature since Dirac~\cite{P.A.M.Dirac-1927}, which discuss the
self-adjoint operator corresponding to the phase difference.
Although there are many theoretical arguments on the self-adjoint
operator which corresponds to the phase difference, a huge number of
experiments to measure the photon phase difference have also been
carried out.
The most recent impressive example is the gravitational-wave detector
which finally succeed direct observations of gravitational wave
through the measurement of the photon phase
difference~\cite{LIGO-GW150914-2016}.
The fundamental principle of the interferometric gravitational-wave
detectors is to imprint gravitational-wave signals to the phase
difference of photons which propagate different paths and to measure
the phase difference between them.


Current gravitational-wave detectors use the DC readout scheme, in
which the output photon power is directly measured.
On the other hand, {\it homodyne detections} are regarded as one of
candidates of the readout scheme in the future gravitational-wave
detectors.
From the proposal by Vyatchanin, Matsko, and
Zubova~\cite{S.P.Vyatchanin-A.B.Matsko-1993} in 1993, it has been
believed in the gravitational-wave community that  ``{\it we can
  measure the output quadrature $\hat{b}_{\theta}$ defined by
  \begin{eqnarray}
    \label{eq:hatbthehta-def}
    \hat{b}_{\theta}
    :=
    \cos\theta \hat{b}_{1} + \sin\theta \hat{b}_{2}
  \end{eqnarray}
  by the balanced homodyne
  detection}''~\cite{H.J.Kimble-Y.Levin-A.B.Matsko-K.S.Thorne-S.P.Vyatchanin-2001},
where $\theta$ is the homodyne angle and $\hat{b}_{1,2}$ are the
amplitude and phase quadratures in the two-photon
formulation~\cite{C.M.Caves-B.L.Schumaker-1985}, respectively.
Note that these operators $\hat{b}_{1,2}$ are defined as linear
combinations of the annihilation and creation operators for the output
optical field from the interferometer.
In the case of the interferometric gravitational-wave detectors, the
output quadrature $\hat{b}_{\theta}$ includes gravitational-wave
signal.
Apart from the leakage of the classical carrier field, we can formally
write this output quadrature as
\begin{eqnarray}
  \label{eq:hatbthehta-response}
  \hat{b}_{\theta}(\Omega)
  =
  R(\Omega,\theta) \left(
  \hat{h}_{n}(\Omega,\theta) + h(\Omega)
  \right)
  ,
\end{eqnarray}
where $h(\Omega)$ is a classical gravitational-wave signal in the
frequency domain, $\hat{h}_{n}(\Omega)$ is the noise operator which is
given by the linear combination of the photon annihilation and
creation operators for the optical fields which are injected to the
main interferometer.
Furthermore, it was pointed out that if we can prepare an appropriate
frequency-dependent homodyne angle $\theta=\theta(\Omega)$, we can
reduce the quantum noise produced by the noise operator
$\hat{h}_{n}(\Omega,\theta)$~\cite{H.J.Kimble-Y.Levin-A.B.Matsko-K.S.Thorne-S.P.Vyatchanin-2001}.
Therefore, in interferometric gravitational-wave detectors, it is
important to develop technique of the homodyne detection,
theoretically and experimentally, for the extraction of the
information of the gravitational-wave signal $h(\Omega)$ through the
measurement of some expectation values of some quantum operators which
are related to the operator $\hat{b}_{\theta}$.


On the other hand, in quantum measurement theory, homodyne detections
are known as the measurement scheme of a linear combination of the
photon annihilation and creation
operators~\cite{H.M.Wiseman-G.J.Milburn-book-2009}.
As noted above, the operator $\hat{b}_{\theta}$ is constructed from
linear combinations of the photon annihilation and creation operators
$\hat{b}$ and $\hat{b}^{\dagger}$.
This means that if we can measure the both expectation values of the
photon annihilation operator $\hat{b}$ and the creation operator
$\hat{b}^{\dagger}$ themselves, we can calculate the expectation value
of the operator $\hat{b}_{\theta}$ from these expectation values.


In this Letter, we report our rediscovery of the eight-port homodyne
detection~\cite{N.G.Walker-J.E.Carrol-1986} as the measurements
of the expectation values of the photon annihilation and creation
operator themselves.
We explicitly show that we can calculate the expectation value of the
operator $\hat{b}_{\theta}$ through the expectation values of the
photon numbers of the output from the eight-port homodyne detection.
We also briefly discuss the noise spectral density in the case where
we apply our results to a readout scheme of interferometric
gravitational-wave detectors.


\section{Balanced homodyne detections in the Heisenberg picture}
\label{sec:Conv_Homodyne_Heisenberg}

\begin{figure}[ht]
  \centering
  \includegraphics[width=0.45\textwidth]{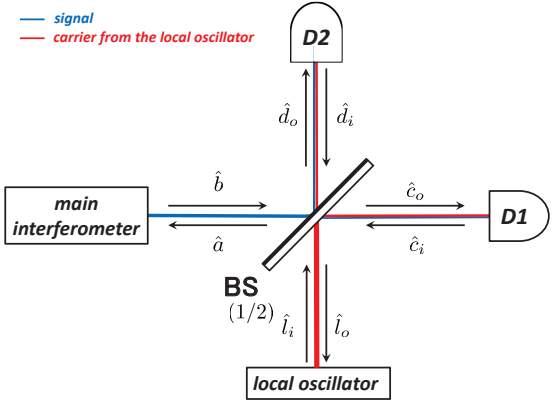}
  \caption{
    Configuration of the interferometer for the balanced homodyne
    detection.
    The beam splitter BS is 50:50.
    The notations of the quadratures $\hat{a}$, $\hat{b}$,
    $\hat{c}_{o}$, $\hat{c}_{i}$, $\hat{d}_{o}$, $\hat{d}_{i}$,
    $\hat{l}_{o}$, and $\hat{l}_{i}$ are also given in this figure.
  }
  \label{fig:Balanced_homodyne_detection}
\end{figure}

First, we briefly review a quantum mechanical description of the
balanced homodyne
detection~\cite{H.M.Wiseman-G.J.Milburn-book-2009}
in the Heisenberg picture depicted in Fig.~\ref{fig:Balanced_homodyne_detection}.


As well known, in interferometers, the electric field operator
associated with the annihilation operator $\hat{a}(\omega)$ at time
$t$ and the length of the propagation direction $z$ in interferometers
is described by
\begin{eqnarray}
  \hat{E}_{a}(t-z)
  &=&
      \int_{-\infty}^{+\infty}
      \frac{d\omega}{2\pi}
      \sqrt{\frac{2\pi\hbar|\omega|}{{\cal A}c}}
      e^{-i\omega(t-z)}
      \nonumber\\
  && \quad\quad
     \times
     \left\{
     \hat{a}(\omega) \Theta(\omega)
     +
     \hat{a}^{\dagger}(-\omega) \Theta(-\omega)
     \right\}
     ,
     \label{eq:electric_field_kouchan_notation}
\end{eqnarray}
where ${\cal A}$ is the cross-sectional area of the optical beam,
$\Theta(\omega)$ is the Heaviside step function, and the annihilation
operator $\hat{a}(\omega)$ satisfies the usual commutation relation
$\left[\hat{a}(\omega),\hat{a}^{\dagger}(\omega)\right]$ $=$ $2 \pi
\delta(\omega-\omega')$.
Throughout this letter, we denote the quadrature $\hat{a}$ as that for
the input to the main interferometer.
On the other hand, we denote the output quadrature from the main
interferometer by $\hat{b}$.
Furthermore, in the homodyne detections, we use the electric field
whose state is a coherent state, which comes from the {\it local
  oscillator}.
The quadrature associated with the electric field from the local
oscillator is denoted by $\hat{l}_{i}$ and the state for the
quadrature $\hat{l}_{i}$ is the coherent state
$|\gamma\rangle_{l_{i}}$ which satisfies
\begin{eqnarray}
  \label{eq:hatli-coherent-state-def}
  \hat{l}_{i}(\omega) |\gamma\rangle_{l_{i}}
  =
  \gamma(\omega) |\gamma\rangle_{l_{i}}.
\end{eqnarray}
Here, $\gamma=\gamma(\omega)$ is the complex eigenvalue for the
coherent state $|\gamma\rangle_{l_{i}}$.


Through the notation of the electric fields as
(\ref{eq:electric_field_kouchan_notation}), we consider the balanced
homodyne detection depicted in
Fig.~\ref{fig:Balanced_homodyne_detection}.
We assign the notation of the photon annihilation operators $\hat{a}$,
$\hat{b}$, $\hat{c}_{o}$, $\hat{c}_{i}$, $\hat{d}_{o}$,
$\hat{d}_{i}$, $\hat{l}_{o}$, and $\hat{l}_{i}$ as in
Fig.~\ref{fig:Balanced_homodyne_detection}.
In the balanced homodyne detection, we detect the photon numbers
$\hat{n}_{c_{o}}:=\hat{c}_{o}^{\dagger}\hat{c}_{o}$ and
$\hat{n}_{d_{o}}:=\hat{d}_{o}^{\dagger}\hat{d}_{o}$ through the
photodetectors D1 and D2, respectively.
We also assume the transmissivity of the beam splitter is 50:50.
From the field junction conditions at the beamsplitter in
Fig.~\ref{fig:Balanced_homodyne_detection}, we obtain the relations of
the quadratures $\hat{c}_{o}$, $\hat{d}_{o}$, $\hat{b}$, and
$\hat{l}_{i}$ as
\begin{eqnarray}
  \hat{c}_{o}
  =
  \frac{
  \hat{b} + \hat{l}_{i}
  }{
  \sqrt{2}
  }
  ,\quad
  \hat{d}_{o}
  =
  \frac{
  \hat{b} - \hat{l}_{i}
  }{
  \sqrt{2}
  }
  \label{eq:BHD-BS-junction}
\end{eqnarray}
and the difference of these photon-number expectation values yields
the expectation value of a linear combination of the output quadrature
$\hat{b}$ as
\begin{eqnarray}
  \langle\hat{n}_{c_{o}}\rangle
  -
  \langle\hat{n}_{d_{o}}\rangle
  =
  \left\langle
  \gamma^{*}
  \hat{b}
  +
  \gamma
  \hat{b}^{\dagger}
  \right\rangle
  .
  \label{eq:BHD-output-expression}
\end{eqnarray}
Here, we used the state for the field from the local oscillator is in
the coherent state (\ref{eq:hatli-coherent-state-def}).
This corresponds to the measurement of the expectation value of the
operator $\hat{s}$ defined by
\begin{eqnarray}
  \label{eq:BHD-operator-def}
  \hat{s}
  :=
  \hat{n}_{c_{o}}
  -
  \hat{n}_{d_{o}}
  =
  \hat{l}_{i}^{\dagger} \hat{b}
  +
  \hat{b}^{\dagger} \hat{l}_{i}
  .
\end{eqnarray}
Note that the linear combination (\ref{eq:BHD-output-expression}) does
not directly yield the expectation value of the output field quadrature
$\hat{b}$ itself, but the phase of the right-hand side in
Eq.~(\ref{eq:BHD-output-expression}) yields the cosine of the relative
phase between the output field and the coherent state from the local
oscillator.
From the view point of the measurement of the quadrature
(\ref{eq:hatbthehta-response}), we want to measure both cosine and
sine parts of the phase of the quadrature $\hat{b}$ with a fixed phase
of the coherent state from the local oscillator.
This is accomplished by the measurements of the expectation values of
the photon annihilation and creation operators themselves through the
``eight-port homodyne detection'' discussed in
Refs.~\cite{N.G.Walker-J.E.Carrol-1986}.


\section{Expectation values of photon annihilation and creation operators}
\label{sec:Expectation_values_of_photon_annihilation_and_creation_operators}

\begin{figure}[ht]
  \centering
  \includegraphics[width=0.48\textwidth]{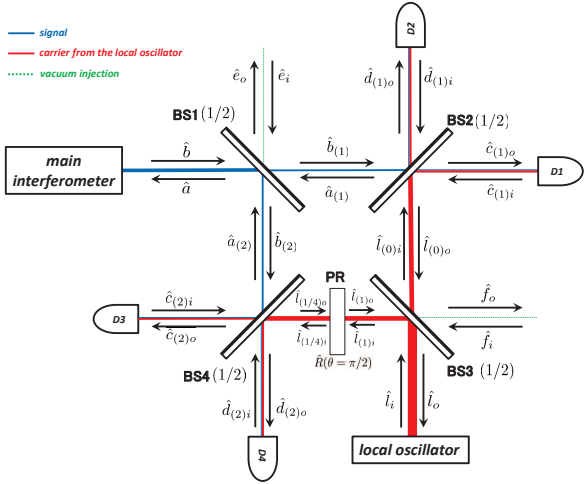}
  \caption{
    A realization of the measurement process of the expectation values
    of the signal photon annihilation operator $\hat{b}$ and creation
    operator $\hat{b}^{\dagger}$ themselves.
    In this figure, ``BS'' is the beam splitter, ``PR'' is the
    phase rotator.
    We assume that all beamsplitters are 50:50.
    To carry out two balanced homodyne detections, we separate the
    signal photon field associated with the quadrature $\hat{b}$ from
    the main interferometer and the photon field associated with the
    quadrature $\hat{l}_{i}$ from the local oscillator through the
    beam splitters BS1 and BS3, respectively.
    One of these two paths is used for the usual balanced homodyne
    detection through the beam splitter BS2 and the photodetectors D1
    and D2.
    We introduce PR on the path between BS3 and BS4 to add
    $\pi/2$-phase offset to the coherent state from the local
    oscillator and we perform the usual balanced homodyne detection
    after this phase addition through the beam splitter B4 and the
    photodetector D3 and D2.
    The notation of the quadratures for the photon fields are also
    described in this figure.
  }
  \label{fig:DoubleBalancedHomodyneDetection-configuration}
\end{figure}

The interferometer configuration of the eight-port homodyne detection
is depicted in
Fig.~\ref{fig:DoubleBalancedHomodyneDetection-configuration}.
Here, we assume that we already knew both of the amplitude and the
phase of the complex amplitude $\gamma$ for the coherent state from
the local oscillator.
We also assume that all beam splitters in
Fig.~\ref{fig:DoubleBalancedHomodyneDetection-configuration} are
50:50 in this letter.


As depicted in
Fig.~\ref{fig:DoubleBalancedHomodyneDetection-configuration},
at the beam splitter 1 (BS1), the output signal $\hat{b}$ from the
main interferometer is separated into two parts, which we denote
$\hat{b}_{(1)}$ and $\hat{b}_{(2)}$, respectively.
In addition to the output quadrature $\hat{b}$, the additional
noise source may be inserted, whose quadrature is denoted by
$\hat{e}_{i}$ and assume that the state for $\hat{e}_{i}$ is the
vacuum state.
Then, the junction conditions for the quadratures at BS1 yield
\begin{eqnarray}
  \label{eq:DBHD-BS1-junction-hatb(1)-hatb(2)}
  \hat{b}_{(1)}
  =
  \frac{
  \hat{b} - \hat{e}_{i}
  }{
  \sqrt{2}
  }
  ,
  \quad
  \hat{b}_{(2)}
  =
  \frac{
  \hat{b} + \hat{e}_{i}
  }{
  \sqrt{2}
  }
  .
\end{eqnarray}


On the other hand, at the beam splitter 3 (BS3), the incident electric
field from the local oscillator is in the coherent state and its
quadrature is denoted by $\hat{l}_{i}$.
Further, from the configuration depicted in
Fig.~\ref{fig:DoubleBalancedHomodyneDetection-configuration}, another
incident field to BS3 should be taken into account.
We denote the quadrature for this additional field as $\hat{f}_{i}$
and assume that the state for $\hat{f}_{i}$ is the vacuum state.
The beam splitter BS3 separate the electric field into two paths.
We denote the quadrature associated with this electric field which
goes from BS3 to BS2 by $\hat{l}_{(0)i}$.
The electric field along another path from BS3 is towards the beam
splitter 4 (BS4) and we denote the quadrature for this field by
$\hat{l}_{(1)i}$.
By the beam splitter condition, quadratures $\hat{l}_{(0)i}$ and
$\hat{l}_{(1)i}$ are determined by the equation
\begin{eqnarray}
  \label{eq:hatl(0)i-hatl(1)ihatli-hatfi-rel}
  \hat{l}_{(0)i}
  =
  \frac{
  \hat{l}_{i}-\hat{f}_{i}
  }{
  \sqrt{2}
  }
  , \quad
  \hat{l}_{(1)i}
  =
  \frac{
  \hat{l}_{i}+\hat{f}_{i}
  }{
  \sqrt{2}
  }
  .
\end{eqnarray}


The field associated with the quadrature $\hat{l}_{(0)i}$ is
used the balanced homodyne detection through the beam splitter 2
(BS2).
On the other hand, the field associated with the quadrature
$\hat{l}_{(1)i}$ is used the balanced homodyne detection through the
beam splitter 4 (BS4) after introducing the phase offset $\pi/2$.
This phase offset is introduced by the phase rotator (PR) between BS3
and BS4.
Due to this phase rotator, the quadrature $\hat{l}_{(1)i}$ is changed
into the quadrature $\hat{l}_{(1/4)i}$ as
\begin{eqnarray}
  \label{eq:phase_rotation_operator_operation}
  \hat{l}_{(1/4)i}
  =
  i
  \hat{l}_{(1)i}
  .
\end{eqnarray}
This quadurature $\hat{l}_{(1/4)i}$ is directly used the balanced
homodyne detection through BS4.


In the balanced homodyne detection through the beam splitter BS2, the
output quadratures  $\hat{c}_{(1)o}$ and $\hat{d}_{(1)o}$ are
related to the input quadratures $\hat{b}_{(1)}$ and $\hat{l}_{(0)i}$
as
\begin{eqnarray}
  \hat{c}_{(1)o}
  =
  \frac{
  \hat{b}_{(1)}
  +
  \hat{l}_{(0)i}
  }{
  \sqrt{2}
  }
  ,
  \quad
  \hat{d}_{(1)o}
  =
  \frac{
  \hat{l}_{(0)i}
  -
  \hat{b}_{(1)}
  }{
  \sqrt{2}
  }
  .
  \label{eq:DBHD-D1-D2-output}
\end{eqnarray}
The photon numbers
$\hat{n}_{c_{(1)o}}:=\hat{c}_{(1)o}^{\dagger}\hat{c}_{(1)o}$ and $\hat{n}_{d_{(1)o}}:=\hat{d}_{(1)o}^{\dagger}\hat{d}_{(1)o}$
of the output fields associated with the
quadratures $\hat{c}_{(1)o}$ and $\hat{d}_{(1)o}$ are detected
through the photodetector D1 and D2 in
Fig.~\ref{fig:DoubleBalancedHomodyneDetection-configuration},
respectively.
These photon numbers are given in terms of the quadrature $\hat{b}$,
$\hat{l}_{i}$, $\hat{e}_{i}$, and $\hat{f}_{i}$ through
Eqs.~(\ref{eq:DBHD-BS1-junction-hatb(1)-hatb(2)}),
(\ref{eq:hatl(0)i-hatl(1)ihatli-hatfi-rel}), and
(\ref{eq:DBHD-D1-D2-output}).
The balanced homodyne detection from the photodetector D1 and D2
yields the expectation value
\begin{eqnarray}
  \label{eq:BHD-D1-D2-result}
  2
  \left(
  \langle\hat{n}_{c_{(1)o}}\rangle
  -
  \langle\hat{n}_{d_{(1)o}}\rangle
  \right)
  =
  \left\langle
  \gamma^{*}
  \hat{b}
  +
  \gamma
  \hat{b}^{\dagger}
  \right\rangle
  ,
\end{eqnarray}
as Eq.~(\ref{eq:BHD-output-expression}), which is also regarded as the
expectation value of the operator
\begin{eqnarray}
  \label{eq:hats-D1D2-operator-original-def}
  \hat{s}_{D1D2}
  &:=&
       2
       \left(
       \hat{n}_{c_{(1)o}}
       -
       \hat{n}_{d_{(1)o}}
       \right)
       \\
  &=&
      \hat{b} \hat{l}_{i}^{\dagger}
      + \hat{b}^{\dagger} \hat{l}_{i}
      \nonumber\\
  &&
     - \hat{b} \hat{f}_{i}^{\dagger}
     - \hat{b}^{\dagger} \hat{f}_{i}
     + \hat{e}_{i} \hat{f}_{i}^{\dagger}
     + \hat{e}_{i}^{\dagger} \hat{f}_{i}
     - \hat{e}_{i} \hat{l}_{i}^{\dagger}
     - \hat{e}_{i}^{\dagger} \hat{l}_{i}
     .
     \label{eq:hats-D1D2-operator-expression-2}
\end{eqnarray}
In the right-hand side of
Eq.~(\ref{eq:hats-D1D2-operator-expression-2}), the first line gives
Eq.~(\ref{eq:BHD-D1-D2-result}), the second line is the vacuum
contributions.


Similarly, the balanced homodyne detection through the beam splitter
BS4 and photodetectors D3 and D4 yields the expectation value
\begin{eqnarray}
  \label{eq:BHD-D3-D4-result}
  2i
  \left(
  \langle\hat{n}_{d_{(2)o}}\rangle
  -
  \langle\hat{n}_{c_{(2)o}}\rangle
  \right)
  =
  \left\langle
  \gamma^{*}
  \hat{b}
  -
  \gamma
  \hat{b}^{\dagger}
  \right\rangle
  ,
\end{eqnarray}
which is regarded as the expectation value of the operator
\begin{eqnarray}
  \label{eq:hats-D3D4-operator-original-def}
  \hat{s}_{D3D4}
  &:=&
       2i
       \left(
       \hat{n}_{d_{(2)o}}
       -
       \hat{n}_{c_{(2)o}}
       \right)
  \\
  &=&
      \hat{l}_{i}^{\dagger}
      \hat{b}
      -
      \hat{b}^{\dagger}
      \hat{l}_{i}
      \nonumber\\
  &&
     - \hat{b}^{\dagger} \hat{f}_{i}
     + \hat{b} \hat{f}_{i}^{\dagger}
     + \hat{l}_{i}^{\dagger} \hat{e}_{i}
     + \hat{f}_{i}^{\dagger} \hat{e}_{i}
     - \hat{e}_{i}^{\dagger} \hat{l}_{i}
     - \hat{e}_{i}^{\dagger} \hat{f}_{i}
     .
  \label{eq:hats-D3D4-operator-expression-2}
\end{eqnarray}
In the right-hand side of
Eq.~(\ref{eq:hats-D3D4-operator-expression-2}), the first line gives
Eq.~(\ref{eq:BHD-D3-D4-result}) and the second line is the vacuum
contributions.
Here, we have to emphasize that the overall factors in
Eqs.~(\ref{eq:BHD-D3-D4-result}) and
(\ref{eq:hats-D3D4-operator-expression-2}) are purely imaginary which
break the self-adjointness of our result.


Since we assumed that we already knew the complex amplitude $\gamma$,
from Eqs.~(\ref{eq:BHD-D1-D2-result}) and (\ref{eq:BHD-D3-D4-result}),
we can calculate the expectation values of operators $\hat{b}$ and
$\hat{b}^{\dagger}$ as
\begin{eqnarray}
  &&
     \label{eq:expecation_valu_of_hatb-result}
     \frac{1}{2\gamma^{*}}
     \left(
     \left\langle
     \hat{s}_{D1D2}
     \right\rangle
     +
     \left\langle
     \hat{s}_{D3D4}
     \right\rangle
     \right)
     =
     \left\langle\hat{b}\right\rangle
     ,
  \\
  &&
     \label{eq:expecation_valu_of_hatbdagger-result}
     \frac{1}{2\gamma}
     \left(
     \left\langle
     \hat{s}_{D1D2}
     \right\rangle
     -
     \left\langle
     \hat{s}_{D3D4}
     \right\rangle
     \right)
     =
     \left\langle\hat{b}^{\dagger}\right\rangle
     .
\end{eqnarray}
Here, we have to emphasize that the expectation value of the operators
$\hat{s}_{D1D2}$ and $\hat{s}_{D3D4}$ are given through the
measurement of the expectation values of photon numbers at  the
photodetector D1, D2, D3, D4, and the complex amplitude $\gamma$ for
the coherent state from the local oscillator.
Similar formulae were also derived in
Refs.~\cite{E.Shchukin-T.Richter-W.Vogel-2005} in the context of the
characterization of the nonclassicality of the system.


The  noise spectral density $S_{\hat{Q}}(\omega)$ defined by
\begin{eqnarray}
  \frac{1}{2} S_{\hat{Q}}(\omega) 2 \pi \delta(\omega-\omega')
  :=
  \frac{1}{2} \left\langle
  \hat{Q}\hat{Q}^{'\dagger}
  +
  \hat{Q}^{'\dagger}\hat{Q}
  \right\rangle
  ,
  \label{eq:spectral_density_def}
\end{eqnarray}
for the operator $\hat{Q}$ with its expectation value
$\langle\hat{Q}\rangle=0$ is commonly used in the gravitational-wave
community to evaluate quantum fluctuations in the measurement of the
operator
$\hat{Q}$~\cite{H.J.Kimble-Y.Levin-A.B.Matsko-K.S.Thorne-S.P.Vyatchanin-2001,H.Miao-PhDthesis-2010}.
Here, $\hat{Q}=\hat{Q}(\omega)$ and
$\hat{Q}^{'\dagger}=\hat{Q}^{\dagger}(\omega')$.
In our case, we define the operators
\begin{eqnarray}
  \hat{t}_{b+}
  &:=&
      \frac{1}{2\gamma^{*}}
      \left(
      \hat{s}_{D1D2}
      +
      \hat{s}_{D3D4}
      \right)
      \nonumber\\
  &=&
      \hat{b} \frac{\hat{l}_{i}^{\dagger}}{\gamma^{*}}
     +
     \frac{1}{\gamma^{*}}
     \left(
     - \hat{b}^{\dagger} \hat{f}_{i}
     + \hat{e}_{i} \hat{f}_{i}^{\dagger}
     - \hat{e}_{i}^{\dagger} \hat{l}_{i}
     \right)
     \label{eq:hatb-measure-operator-result}
      ,
  \\
  \hat{t}_{b-}
  &:=&
       \frac{1}{2\gamma}
       \left(
       \hat{s}_{D1D2}
       -
       \hat{s}_{D3D4}
       \right)
       =
      \hat{t}_{b+}^{\dagger}
     .
     \label{eq:hatbdagger-measure-operator-result}
\end{eqnarray}
Eqs.~(\ref{eq:expecation_valu_of_hatb-result}) and
(\ref{eq:expecation_valu_of_hatbdagger-result}) are regarded as the
expectation values of these operators $\hat{t}_{b+}$ and
$\hat{t}_{b-}$, respectively.
We also define the noise operators $\hat{t}_{b+}^{(n)}$,
$\hat{t}_{b-}^{(n)}$, and $\hat{b}^{(n)}$ by
\begin{eqnarray}
  \label{eq:hattbplus-noise-operators}
  \hat{t}_{b+}
  &=:&
       \langle\hat{b}\rangle + \hat{t}_{b+}^{(n)},
       \quad
       \langle\hat{t}_{b+}^{(n)}\rangle=0,
  \\
  \label{eq:hattbminus-noise-operators}
  \hat{t}_{b-}
  &=:&
       \langle\hat{b}^{\dagger}\rangle + \hat{t}_{b-}^{(n)},
       \quad
       \langle\hat{t}_{b-}^{(n)}\rangle=0,
  \\
  \label{eq:hatb-noise-operators}
  \hat{b}
  &=:&
       \langle\hat{b}\rangle + \hat{b}^{(n)},
       \quad
       \langle\hat{b}^{(n)}\rangle=0
       .
\end{eqnarray}


Further, from the interferometer setup in
Fig.~\ref{fig:DoubleBalancedHomodyneDetection-configuration}, we
should regard that the commutators
$\left[\hat{e}_{i},\hat{f}_{i}\right]$,
$\left[\hat{e}_{i},\hat{f}_{i}^{\dagger}\right]$,
$\left[\hat{f}_{i},\hat{l}_{i}\right]$,
$\left[\hat{f}_{i},\hat{l}_{i}^{\dagger}\right]$,
$\left[\hat{l}_{i},\hat{e}_{i}\right]$, and
$\left[\hat{l}_{i},\hat{e}_{i}^{\dagger}\right]$ vanish.
Moreover, we can easily check that the commutators
$\left[\hat{b}, \hat{e}_{i}\right]$,
$\left[\hat{b},\hat{e}_{i}^{\dagger}\right]$,
$\left[\hat{b},\hat{f}_{i}\right]$,
$\left[\hat{b},\hat{f}_{i}^{\dagger}\right]$,
$\left[\hat{b},\hat{l}_{i}\right]$, and
$\left[\hat{b},\hat{l}_{i}^{\dagger}\right]$ also vanish even if the
output quadrature $\hat{b}$ depends on the input quadrature
$\hat{a}$.


Equations~(\ref{eq:hatb-measure-operator-result})--(\ref{eq:hatb-noise-operators})
and the above commutation relations lead us to the noise spectral
densities
\begin{eqnarray}
  S_{\hat{t}_{b+}^{(n)}}(\omega)
  =
  S_{\hat{t}_{b-}^{(n)}}(\omega)
  =
  S_{\hat{b}^{(n)}}(\omega)
  +
  \frac{2}{|\gamma|^{2}}
  \langle
  \hat{n}_{b}
  \rangle
  +
  1
  .
  \label{eq:spectral-density-relation-in-DBHD-eta=1/2}
\end{eqnarray}
This noise spectral density indicates that in addition to the noise
spectral density $S_{\hat{b}^{(n)}}(\omega)$, we have the additional
fluctuations in our measurement of $\langle\hat{b}\rangle$ through the
measurement of the operator $\hat{t}_{b+}$.
We note that the term $2\langle\hat{n}_{b}\rangle/|\gamma^{2}|$ will be
negligible if $\langle\hat{n}_{b}\rangle \ll |\gamma^{2}|$.
On the other hand, the the last term $1$ in
Eq.~(\ref{eq:spectral-density-relation-in-DBHD-eta=1/2}), which comes
from the shot noise from the additional input vacuum fields, is not
controllable.


\section{Expectation value of the operator $\hat{b}_{\theta}$ and its noise}
\label{sec:hatbtheta_expectation_value_and_its_noise}


In the case of gravitational-wave detectors, the two-photon
formulation~\cite{C.M.Caves-B.L.Schumaker-1985} is always used.
In this formulation, we consider the sideband fluctuations around the
classical carrier field which proportional to $\cos\omega_{0}t$.
The sideband fluctuations are described by the quadrature
$\hat{b}_{\pm}(\Omega):=\hat{b}(\omega_{0}\pm\Omega)$ and we introduce
quadratures $\hat{b}_{1,2}$ by
\begin{eqnarray}
  \label{eq:amplitude-phase-quadratures-defs}
  \hat{b}_{1}
  :=
  \frac{1}{\sqrt{2}} \left(\hat{b}_{+}+\hat{b}_{-}^{\dagger}\right)
  , \quad
  \hat{b}_{2}
  :=
  \frac{1}{\sqrt{2}i} \left(\hat{b}_{+}-\hat{b}_{-}^{\dagger}\right)
  .
\end{eqnarray}
These are the definitions of the amplitude (phase) quadrature
$\hat{b}_{1}$ ($\hat{b}_{2}$) in Eq.~(\ref{eq:hatbthehta-def}).
We also consider the sideband quadratures of the local oscillator
$\hat{l}_{i\pm}=\hat{l}_{i}(\omega_{0}\pm\Omega)$ and assume
that the state of the local oscillator is the coherent state
(\ref{eq:hatli-coherent-state-def}) in which the eigenvalues of
operators $\hat{l}_{i\pm}$ are given by
$\gamma_{\pm}:=\gamma(\omega_{0}\pm\Omega)$, respectively.


In this situation, we can obtain the information of the expectation
values $\langle\hat{s}_{\pm}\rangle$ $=$
$\left\langle
  \gamma_{\pm}^{*}\hat{b}_{\pm} + \gamma_{\pm}\hat{b}_{\pm}^{\dagger}
\right\rangle$ through the usual balanced homodyne detection, where we
used the definition (\ref{eq:BHD-operator-def}) and
$\hat{s}_{\pm}:=\hat{s}(\omega_{0}\pm\Omega)$.
Within the linear level of quadratures, we can obtain the expectation
value of any linear combination
\begin{eqnarray}
  && \!\!\!\!\!\!\!\!\!\!\!\!\!\!\!\!
     \sqrt{2}\left(
     \alpha \left\langle\hat{s}_{+}\right\rangle
     +
     \beta \left\langle\hat{s}_{-}\right\rangle
     \right)
     \nonumber\\
  &=&
      \left(
      \alpha
      \gamma_{+}^{*}
      +
      \beta
      \gamma_{-}
      \right)
      \langle
      \hat{b}_{1}
      \rangle
      +
      i
      \left(
      \alpha
      \gamma_{+}^{*}
      -
      \beta
      \gamma_{-}
      \right)
      \langle
      \hat{b}_{2}
      \rangle
      \nonumber\\
  &&
      +
      \left(
      \alpha
      \gamma_{+}
      +
      \beta
      \gamma_{-}^{*}
      \right)
     \langle
      \hat{b}_{1}^{\dagger}
     \rangle
      +
      i
      \left(
      -
      \alpha
      \gamma_{+}
      +
      \beta
      \gamma_{-}^{*}
      \right)
     \langle
      \hat{b}_{2}^{\dagger}
     \rangle
      .
  \label{eq:alphahatsplus+betahatsminus-exp-value}
\end{eqnarray}
with complex coefficients $\alpha$ and $\beta$.
The problem whether or not we can measure the expectation value of the
operator $\hat{b}_{\theta}$ defined by (\ref{eq:hatbthehta-def})
through the conventional balanced homodyne detection is reduced to the
problem whether or not the linear combination
(\ref{eq:alphahatsplus+betahatsminus-exp-value}) gives the linear
combination of the expectation values $\langle\hat{b}_{1}\rangle$ and
$\langle\hat{b}_{2}\rangle$ through an appropriate choice of $\alpha$,
$\beta$, and $\gamma_{\pm}$.
Within the linear optics,
Eq.~(\ref{eq:alphahatsplus+betahatsminus-exp-value}) yields the
expectation value of the linear combination of
$\langle\hat{b}_{1}\rangle$ and $\langle\hat{b}_{2}\rangle$ if and
only if there exists a nontrivial solution with
$\alpha\beta\gamma_{+}\gamma_{-}\neq 0$ of the matrix equation
\begin{eqnarray}
  \label{eq:hatb1-hatb2-case-condition}
  \left(
    \begin{array}{cc}
        \gamma_{+} & \gamma_{-}^{*} \\
      - \gamma_{+} & \gamma_{-}^{*}
    \end{array}
  \right)
  \left(
    \begin{array}{c}
      \alpha \\
      \beta
    \end{array}
  \right)
  =
  \left(
    \begin{array}{c}
      0 \\
      0
    \end{array}
  \right)
  .
\end{eqnarray}
However, we can easily show that
Eq.~(\ref{eq:hatb1-hatb2-case-condition}) has no nontrivial solution
satisfies the condition $\alpha\beta\gamma_{+}\gamma_{-}\neq 0$.
This means that any choice of $\alpha$, $\beta$ and $\gamma_{\pm}$
never yields the linear combination (\ref{eq:hatbthehta-def}).
Thus, we cannot measure the expectation value of the operator
$\hat{b}_{\theta}$ through the conventional balanced homodyne
detection depicted in
Fig.~\ref{fig:Balanced_homodyne_detection}~\cite{K.Nakamura-M.-K.Fujimoto-2017-DBHD-full-paper}.


On the other hand, the measurement of the expectation value of
$\hat{b}_{\theta}$ is possible through the eight-port homodyne
detection depicted in
Fig.~\ref{fig:DoubleBalancedHomodyneDetection-configuration}.
In this interferometer setup, we obtain the expectation values of the
operators $\hat{s}_{D1D2}$ and $\hat{s}_{D3D4}$ defined by
Eqs.~(\ref{eq:hats-D1D2-operator-original-def}) and
(\ref{eq:hats-D3D4-operator-original-def}), respectively.
Since we consider the sideband fluctuations around the classical
carrier with the frequency $\omega_{0}$, we have four expectation
values
\begin{eqnarray}
  \label{eq:hatsD1D2pm-exp-value}
  \langle\hat{s}_{D1D2\pm}\rangle := \langle\hat{s}_{D1D2}(\omega_{0}\pm\Omega)\rangle
  =
  \left\langle
  \gamma_{\pm}^{*}\hat{b}_{\pm} + \gamma_{\pm}\hat{b}_{\pm}^{\dagger}
  \right\rangle
  , \\
  \label{eq:hatsD3D4pm-exp-value}
  \langle\hat{s}_{D3D4\pm}\rangle := \langle\hat{s}_{D3D4}(\omega_{0}\pm\Omega)\rangle
  =
  \left\langle
  \gamma_{\pm}^{*}\hat{b}_{\pm} - \gamma_{\pm}\hat{b}_{\pm}^{\dagger}
  \right\rangle
  .
\end{eqnarray}
Inspecting the expectation values (\ref{eq:hatsD1D2pm-exp-value}) and
(\ref{eq:hatsD3D4pm-exp-value}), we first consider the operators
\begin{eqnarray}
  \label{eq:hattD1D2+}
  \hat{t}_{D1D2+}
  &:=&
       \frac{1}{\sqrt{2}} \left(
       \frac{\hat{s}_{D1D2+}}{|\gamma_{+}|}
       +
       \frac{\hat{s}_{D1D2-}}{|\gamma_{-}|}
       \right)
       ,
  \\
  \label{eq:hattD3D4-}
  \hat{t}_{D3D4-}
  &:=&
       \frac{1}{\sqrt{2}}
       \left(
       \frac{\hat{s}_{D3D4+}}{|\gamma_{+}|}
       -
       \frac{\hat{s}_{D3D4-}}{|\gamma_{-}|}
       \right)
       ,
\end{eqnarray}
where $\gamma_{\pm}=:|\gamma_{\pm}|e^{i\theta_{\pm}}$.
Here, we choose the phase $\theta_{\pm}$ of the coherent amplitude
$\gamma_{\pm}$ so that  $\theta_{\pm}=\theta$.
Furthermore, we assume that $|\gamma_{\pm}|=|\gamma|$, for
simplicity.
Due to this phase choice, we easily obtain the expectation value of
the operator $\hat{t}_{\theta}:=(\hat{t}_{D1D2+}+\hat{t}_{D3D4-})/2$ as
\begin{eqnarray}
  \langle
  \hat{t}_{\theta}
  \rangle
  =
  \left\langle
  \cos\theta\hat{b}_{1} + \sin\theta\hat{b}_{2}
  \right\rangle
  =
  \langle\hat{b}_{\theta}\rangle
  .
  \label{eq:ahatttheta-expectation-value}
\end{eqnarray}
Thus, we can obtain the expectation value of the operator
$\hat{b}_{\theta}$ as the direct output of the eight-port homodyne
detection.


We have to emphasis  that our measurement of the expectation value of
the operator $\hat{b}_{\theta}$ is an indirect measurement of the
expectation value of operator $\hat{b}_{\theta}$.
To obtain the expectation value of the operator $\hat{b}_{\theta}$, we
just calculate the linear combination of the expectation values of the
output photon number operators with the complex coefficients.
Since we directly measure the photon number operators which are
self-adjoint operators, our analyses and results do not contradict to
the axiom of the quantum theory.
The fact that our measurement of the expectation value
$\hat{b}_{\theta}$ is an indirect measurement directly leads the fact
that the noise in our measurement cannot be given by the expectation
value of the square of the operators $\hat{b}_{\theta}$.
Actually, through the separation
\begin{eqnarray}
  \label{eq:hattthetanoise-hatbthetanoisedef}
  \hat{t}_{\theta}
  :=
  \langle\hat{b}_{\theta}\rangle
  +
  \hat{t}^{(n)}_{\theta}
  ,
  \quad
  \hat{b}_{\theta}
  :=
  \langle\hat{b}_{\theta}\rangle
  +
  \hat{b}^{(n)}_{\theta}
  ,
\end{eqnarray}
we can also evaluate the noise-spectral density of this measurement
through the similar derivation to
Eq.~(\ref{eq:spectral-density-relation-in-DBHD-eta=1/2}) as
\begin{eqnarray}
  \label{eq:hattthetanoise-hatbtheta-noise-relation}
  S_{\hat{t}_{\theta}^{(n)}}(\Omega)
  =
  \hat{S}_{\hat{b}_{\theta}^{(n)}}(\Omega)
  +
  \frac{\langle\hat{n}_{b_{-}}+\hat{n}_{b_{+}}\rangle}{|\gamma|^{2}} + 1.
\end{eqnarray}
This noise spectral density indicates  a noise level in our
measurement of the operator $\hat{b}_{\theta}$.


\section{Gravitational-wave signal referred noise}
\label{sec:gravitational_wave_signal_referred_noise}


The input-output relation of gravitational-wave detectors are given by
Eq.~(\ref{eq:hatbthehta-response}).
Through the above eight-port homodyne detection, we can measure the
expectation value of the operator $\hat{t}_{\theta}$ to obtain the
expectation value $\langle\hat{b}_{\theta}\rangle$ which includes
gravitational-wave signals $h(\Omega)$.
To evaluate the noise spectral density from the input-output relation
(\ref{eq:hatbthehta-response}), we can directly obtain the
signal-referred noise spectral density as
\begin{eqnarray}
  \frac{1}{|R|^{2}} S_{\hat{t}_{\theta}^{(n)}}
  =
  S_{\hat{h}_{(n)}}
  +
  \frac{1}{|R|^{2}}
  \left(
  \frac{\langle\hat{n}_{-}+\hat{n}_{+}\rangle}{|\gamma|^{2}}
  +
  1
  \right)
  .
  \label{eq:signal-referrred-noise-sepctral-desnity}
\end{eqnarray}
This relation between noise-spectral densities yields that the second
term in Eq.~(\ref{eq:signal-referrred-noise-sepctral-desnity}), which
corresponds to the additional noise due to our indirect measurement,
is negligible if the response function $R(\Omega)$ is sufficiently
large.
In this case, the signal-referred noise spectral density
$S_{\hat{t}_{\theta}^{(n)}}/|R|^{2}$ in our indirect measurement
coincides with the signal-referred noise spectral density
$S_{\hat{h}^{(n)}}$.


\section{Summary}
\label{sec:summary}


In summary, we rediscovered  so-called the eight-port homodyne
detection as the device to measure the expectation values of the photon
annihilation and creation operator for the output optical field from
the main interferometer and discuss its application to a readout
scheme of the gravitational-wave detectors.
As we emphasized above, we just proposed an indirect measurement which
yields the expectation value $\langle\hat{b}_{\theta}\rangle$ through
the calculation by the linear combination of four photon-number
expectation values with complex coefficients.
This indirect measurement of the operator $\hat{b}_{\theta}$ leads the
additional fluctuations in the measurement as in
Eq.~(\ref{eq:hattthetanoise-hatbtheta-noise-relation}).


Furthermore, we explained the outline of our proof for the
assertion that {\it we cannot measure the expectation value of the
  operator $\hat{b}_{\theta}$ in Eq.~(\ref{eq:hatbthehta-response}) by
  the balanced homodyne detection, but it is possible by the
  eight-port homodyne detection if the coherent state from the local
  oscillator is appropriately prepared.}
We also evaluated the noise spectral densities of these measurements
and discussed the noise level when we apply our result to the readout
scheme of interferometric gravitational-wave detectors.


We call this measurement scheme to measure the expectation values of
the photon annihilation and creation operators, or operator
$\hat{b}_{\theta}$ defined by Eq.~(\ref{eq:hatbthehta-def}) using the
eight-port homodyne detection as the ``{\it double balanced homodyne
  detection}.''
More detailed analysis of our double balanced homodyne detection will
be explained in
Ref.~\cite{K.Nakamura-M.-K.Fujimoto-2017-DBHD-full-paper}.


\section*{Acknowledgments}


K.N. acknowledges to Dr. Tomotada Akutsu and the other members of the
GWPO in NAOJ for their continuous encouragement to our research and
also appreciate Prof. Akio Hosoya, Prof. Izumi Tsutsui, and Dr. Hiroyuki
Takahashi for their supports and encouragement.



\end{document}